\documentclass[proceedings]{stacs}
\stacsheading{2009}{541--552}{Freiburg}
\firstpageno{541}

\usepackage[latin1]{inputenc}
\usepackage[T1]{fontenc}
\usepackage{amsmath}
\usepackage{amssymb}
\usepackage{amscd}
\usepackage{graphics}
\usepackage{graphicx} 
\usepackage{gastex}
\usepackage[usenames]{color}
\usepackage{subfigure}
\usepackage{wasysym}
\usepackage{xspace}
\usepackage{yfonts}

\newcommand{\Pick}{\mathop{\mathrm{Pick}}}
\newcommand{\Picks}{\mathop{\mathrm{Pick^*}}}

\newcommand{\muller}{\calF}

\newcommand{\mf}{\ensuremath{m_\mathcal{F}}}

\newcommand{\rf}{r_{\!\calF}}

\newcommand{\af}{\ensuremath{\mathcal{A}_\mathcal{F}}}

\newcommand{\am}{\ensuremath{\mathcal{A_F}}}
\newcommand{\ram}{\rf}
\newcommand{\m}{\ensuremath{\mathcal{F}}}

\newcommand{\zfc}{\ensuremath{\mathcal{Z_{F,C}}}}

\newcommand{\event}{P}

\newcommand{\calA}{{\mathcal A}}

\newcommand{\calC}{{\mathcal C}}
\newcommand{\calD}{{\mathcal D}}
\newcommand{\calE}{{\mathcal E}}
\newcommand{\calF}{{\mathcal F}}
\newcommand{\calG}{{\mathcal G}}

\newcommand{\calP}{{\mathcal P}}

\newcommand{\calS}{{\mathcal S}}
\newcommand{\calT}{{\mathcal T}}

\newcommand{\calZ}{{\mathcal Z}}

\newcommand{\sigman}{\sigma^{\tt n}}
\newcommand{\sigmau}{\sigma^{\tt u}}
\newcommand{\re}{r_\calE}
\newcommand{\dfc}{\calD_{\calF,\calC}}
\newcommand{\are}{\calA_\calE}
\renewcommand{\iff}{if and only if\xspace}

\newcommand{\thp}{$2\frac{1}{2}$-player\xspace}

\newcommand{\Inf}{\mathop{\mathrm{Inf}}}
\newcommand{\Attr}{\mathop{\mathrm{Attr}}}

\newcommand{\gothA}{\textgoth{A}}
\newcommand{\gothF}{\textgoth{F}}
\newcommand{\gothG}{\textgoth{G}}

\begin{document}

% \title[short title]{title}
\title{Random Fruits on the Zielonka Tree}

% \author[ref]{Short author}{Author}
\author[cwi]{Florian Horn}{Florian Horn}

\address[cwi]{CWI, Amsterdam, The Netherlands}	%optional
\email{f.horn@cwi.nl}

\thanks{This work was carried out during the tenure of an ERCIM "Alain Bensoussan" Fellowship Programme.} %optional

%% mandatory lists of keywords and classifications:
% \titlecomment{OPTIONAL comment concerning the title, \eg, if a variant
% or an extended abstract of the paper has appeared elsewehere}
%%%%%%%%%%%%%%%%%%%%%%%%%%%%%%%%%%%%%%%%%%%%%%%%%%%%%%%%%%%%%%%%%%%%%%%%%%%

%% the abstract has to PRECEDE the command \maketitle:
%% be sure not to issue the \maketitle command twice!

\begin{abstract}
  \noindent Stochastic games are a natural model for the synthesis of controllers confronted to adversarial and/or random actions. In particular, $\omega$-regular games of infinite length can represent reactive systems which are not expected to reach a correct state, but rather to handle a continuous stream of events. One critical resource in such applications is the memory used by the controller. In this paper, we study the amount of memory that can be saved through the use of randomisation in strategies, and present matching upper and lower bounds for stochastic Muller games.
\end{abstract}

\maketitle

%% start the paper here:
\section{Introduction}

A stochastic game arena is a directed graph with three kinds of states: Eve's, Adam's and random states. A token circulates on this arena: when it is in one of Eve's states, she chooses its next location among the successors of the current state; when it is in one of Adam's states, he chooses its next location; and when it is in a random state, the next location is chosen according to a fixed probability distribution. The result of playing the game for $\omega$ moves is an infinite path of the graph. A play is winning either for Eve or for Adam, and the ``winner problem'' consists in determining whether one of the players has a winning strategy, from a given initial state. Closely related problems concern the computation of winning strategies, as well as determining the nature of these strategies: pure or randomised, with finite or infinite memory. There has been a long history of using arenas without random states (2-player arenas) for modelling and synthesising reactive processes \cite{BuchiLandweberAMS69,PnueliRosnerPOPL89}: Eve represents the controller, and Adam the environment. Stochastic (2$\frac{1}{2}$-player) arenas \cite{deAlfaro97}, with the addition of random states, can also model uncontrollable actions that happen according to a random law, rather than by choice of an actively hostile environment. The desired behaviour of the system is traditionally represented as an $\omega$-regular winning condition, which naturally expresses the temporal specifications and fairness assumptions of transition systems \cite{MannaPnueli92}. From this point of view, the complexity of the winning strategies is a central question, since they represent possible implementations of the controllers in the synthesis problem. In this paper, we focus on an important normal form of $\omega$-regular conditions, namely \textit{Muller} winning conditions (see \cite{ThomasSTACS95} for a survey).

In the case of 2-player Muller games, a fundamental determinacy result of B\"uchi and Landweber states that, from any initial state, one of the players has a winning strategy \cite{BuchiLandweberAMS69}. Gurevich and Harrington used the \textit{latest appearance record} (LAR) structure of McNaughton to extend this result to strategies with memory factorial in the size of the game \cite{GurevichHarringtonSTOC82}. Zielonka refines the LAR construction into a tree, and derives from it an elegant algorithm to compute the winning regions \cite{ZielonkaTCS98}. An insightful analysis of the Zielonka tree by Dziembowski, Jurdzinski, and Walukiewicz leads to optimal (and asymmetrical) memory bounds for pure (non-randomised) winning strategies \cite{DziembowskiJurdzinskiWalukiewiczLICS97}. Chatterjee extended these bounds to the case of pure strategies over 2$\frac{1}{2}$-player arenas \cite{Cha07-FSTTCS}. However, the lower bound on memory does not hold for \textit{randomised strategies}, even in non-stochastic arenas: Chatterjee, de Alfaro, and Henzinger show that memoryless randomised strategies are enough for to deal with upward-closed winning conditions  \cite{ChatterjeedeAlfaroHenzingerQEST04}. Chatterjee extends this result in \cite{ChatterjeeFOSSACS07}, showing that conditions with non-trivial upward-closed subsets admit randomised strategies with less memory than pure ones.

\noindent \textbf{Our contributions}. The memory bounds of \cite{ChatterjeeFOSSACS07} are not tight in general, even for 2-player arenas. We give here matching upper and lower bounds for any Muller condition $\calF$, in the form of a number $\rf$ computed from the Zielonka tree of $\calF$:

\begin{itemize}
	\item if Eve has a winning strategy in a \thp game $(\calA,\calF)$, she has a randomised winning strategy with memory $\rf$ (Theorem~\ref{theorem:upper});
	\item there is a 2-player game $(\af,\calF)$ where any randomised winning strategy for Eve has at least $\rf$ memory states (Theorem~\ref{theorem:lower}).
\end{itemize}
Furthermore, the witness arenas we build in the proof of Theorem~\ref{theorem:lower} are significantly smaller than in \cite{DziembowskiJurdzinskiWalukiewiczLICS97}, even though the problem of \textit{polynomial} arenas remains open.

\noindent \textbf{Outline of the paper}. Section~\ref{section:definitions} recalls the classical notions in the area, while Section~\ref{section:former} presents former results on memory bounds and randomised strategies. The next two sections present our main results. In Section~\ref{section:upper}, we introduce the number $\rf$ and show that it is an upper bound on the memory needed to win in any 2$\frac{1}{2}$-game $(\calA,\calF)$. In Section~\ref{section:lower}, we show that this bound is tight. Finally, in Section~\ref{section:conclusions}, we characterise the class of Muller conditions that admit memoryless randomised strategies, and show that for each Muller condition, at least one of the players cannot improve its memory through randomisation.

\section{Definitions}
\label{section:definitions}

We consider turn-based stochastic two-player Muller games. We recall here several classical notions in the field, and refer the reader to \cite{ThomasSTACS95,deAlfaro97} for more details.

\noindent\textbf{Probability Distribution.} A probability distribution $\gamma$ over a set $X$ is a function from $X$ to $[0,1]$ such that $\sum_{x\in X} \gamma(x) = 1$. The set of probability distributions over $X$ is denoted by $\calD(X)$.

\noindent\textbf{Arenas.} A \textit{2$\frac{1}{2}$-player arena} $\mathcal{A}$ over a set of colours $\calC$ consists of a directed finite graph $(\calS,\calT)$, a partition $(\calS_E,\calS_A,\calS_R)$ of $\calS$, a probabilistic transition function $\delta: S_R \rightarrow \mathcal{D}(S)$ such that $\delta(s)(t) > 0 \Leftrightarrow (s,t) \in \calT$, and a partial colouring function $\chi: \calS \rightharpoonup \calC$. The states in $\calS_E$ (resp. $\calS_A$, $\calS_R$) are \textit{Eve's states} (resp. \textit{Adam's states}, \textit{random states}), and are graphically represented as $\fullmoon$'s (resp. $\Box$, $\triangle$). A \textit{2-player arena} is an arena where $\calS_R = \emptyset$. 

A set $U \subseteq \calS$ of states is \textit{$\delta$-closed} if for every random state $u \in U \cap \calS_R$, $(u,t) \in \calT \rightarrow t \in U$. It is \textit{live} if for every non-random state $u \in U \cap (\calS_E \cup \calS_A)$, there is a state $t \in U$ such that $(u,t) \in \calT$. A live and $\delta$-closed subset $U$ induces a \textit{subarena} of $\mathcal{A}$, denoted by $\mathcal{A} \upharpoonright U$.

\noindent\textbf{Plays and Strategies.} An infinite path, or \textit{play}, over the arena $\mathcal{A}$ is an infinite sequence $\rho = \rho_0 \rho_1 \ldots$ of states such that $(\rho_i,\rho_{i+1}) \in \calT$ for all $i \in \mathbb{N}$. The set of states occurring infinitely often in a play $\rho$ is denoted by $\Inf (\rho) = \{s \mid \exists^\infty i \in \mathbb{N}, \rho_i = s\}$. We write $\Omega$ for the set of all plays, and $\Omega_s$ for the set of plays that start from the state $s$.

A \textit{strategy with memory $M$} for Eve on the arena $\calA$ is a (possibly infinite) transducer $\sigma = (M, \sigman, \sigmau)$, where $\sigman$ is the ``next-move'' function from $(\calS_E \times M)$ to $\mathcal{D}(\calS)$ and $\sigmau$ is the ``memory-update'' function, from $(\calS \times M)$ to $\mathcal{D}(M)$. Notice that both the move and the update are randomised: strategies whose memory is deterministic are a different, less compact, model. The strategies for Adam are defined likewise. A strategy $\sigma$ is \textit{pure} if it does not use randomisation. It is \textit{finite-memory} if $M$ is a finite set, and \textit{memoryless} if $M$ is a singleton. Notice that strategies defined in the usual way as functions from $\calS^*$ to $\calS$ can be defined as strategies with infinite memory: the set of memory states is $\calS^*$ and the memory update is $\sigmau(s,w) \mapsto ws$.

Once a starting state $s\in \calS$ and strategies $\sigma \in \Sigma$ for both players are fixed, the outcome of the game is a random walk $\rho_s^{\sigma, \tau}$ for which the probabilities of events are uniquely fixed (an \textit{event} is a measurable set of paths). For an event $\event \in \Omega$, we denote by $\mathbb{P}_s^{\sigma,\tau}(\event)$ the probability that a play belongs to $\event$ if it starts from $s$ and Eve and Adam follow the strategies $\sigma$ and $\tau$. 

A play is consistent with $\sigma$ if for each position $i$ such that $w_i \in \calS_E$, $\mathbb{P}_{w_0}^{\sigma,\tau}(\rho_{i+1} = w_{i+1} \mid \rho_0 = w_0 \ldots \rho_i = w_i) > 0$. The set of plays consistent with $\sigma$ is denoted by $\Omega^\sigma$. Similar notions can be defined for Adam's strategies.

\noindent\textbf{Traps and Attractors.} The \textit{attractor of Eve to the set $U$}, denoted $\Attr_E(U)$, is the set of states where Eve can guarantee that the token reaches the set $U$ with a positive probability. It is defined inductively by:

\begin{center}
$	\begin{array}{cccl}
		\Attr_E^0(U) & = & U & \\
		\Attr_E^{i+1}(U) & = & \Attr_E^{i}(U) & \cup \{s \in \calS_E \cup \calS_R, \exists t \in \Attr_E^{i}(U) \mid (s,t) \in \calT\} \\
		 & & & \cup \{s \in \calS_A \mid \forall t, (s,t) \in E \Rightarrow t \in \Attr_E^{i}(U)\} \\
		\Attr_E(U) & = & \bigcup_{i>0} & \Attr_E^{i}(U) \\
	\end{array}
$
\end{center}

The corresponding \textit{attractor strategy to $U$ for Eve} is a pure and memoryless strategy $a_U$ such that for any state $s \in \calS_E \cap (\Attr_E(U) \setminus U)$, $s \in \Attr_E^{i+1}(U) \Rightarrow a_U(s) \in \Attr_E^{i}(U)$.

The dual notion of \textit{trap for Eve} denotes a set from where Eve cannot escape, unless Adam allows her to do so: a set $U$ is a trap for Eve if and only if $\forall s \in U \cap (\calS_E \cup \calS_R), (s,t) \in \calT \Rightarrow t \in U$ and $\forall s \in U \cap \calS_A, \exists t \in U, (s,t) \in \calT$. Notice that a trap is a ``strong'' notion ---the token can never leave it if Adam does not allow it to do so--- while an attractor is a ``weak'' one ---the token can avoid the target even if Eve uses the attractor strategy. Notice also that a trap (for either player) is always a subarena.

\noindent\textbf{Winning Conditions.} A \textit{winning condition} is a subset $\Phi$ of $\Omega$. A play $\rho$ is \textit{winning for Eve} if $\rho \in \Phi$, and \textit{winning for Adam} otherwise. We consider $\omega$-regular winning conditions formalised as \textit{Muller conditions}. A Muller condition is determined by a subset $\calF$ of the power set $\mathcal{P}(\calC)$ of colours, and Eve wins a play \iff the set of colours visited infinitely often belongs to $\calF$: $ \Phi_{\calF} = \{\rho \in \Omega | \chi(\Inf(\rho)) \in \calF\}$. An example of Muller game is given in Figure~\ref{subfigure:game}. We use it throughout the paper to describe various notions and results.

\begin{figure}[ht]
	\begin{center}
		\subfigure[The game $\gothG = (\gothA,\gothF)$]{
		\label{subfigure:game}
			\unitlength = 2mm
		\begin{picture}(22,17)(0,-2)
%			\put(0,-2){\framebox(22,17){}}
			\gasset{Nw=2,Nh=2}
			\node[Nmr=1](0)(11,14){$c$}
			\node[Nmr=0](1)(1,8){$d$}
			\node[Nmr=0](2)(6,11){$a$}
			\node[Nmr=0](3)(6,5){$b$}
			\node[Nmr=1](4)(11,8){}
			\node[Nmr=0](5)(16,11){}
			\node[Nmr=0](6)(16,5){}
			\node[Nmr=1](7)(21,10){$a$}
			\node[Nmr=1](8)(21,6){$b$}
			\node[Nmr=1](9)(11,2){$c$}
			
			\drawedge(0,4){}
			\drawedge(1,4){}
			\drawedge[curvedepth=-1.3](2,1){}
			\drawedge[curvedepth=1.3](3,1){}
			\drawedge[curvedepth=-1.3](4,2){}
			\drawedge[curvedepth=1.3](4,3){}
			\drawedge[curvedepth=1.3](4,5){}
			\drawedge[curvedepth=-1.3](4,6){}
			\drawedge[curvedepth=-1.3](5,0){}
			\drawedge[curvedepth=1.3](5,7){}
			\drawedge[curvedepth=-1.3](6,8){}
			\drawedge[curvedepth=1.3](6,9){}
			\drawedge[curvedepth=1.3](7,4){}
			\drawedge[curvedepth=-1.3](8,4){}
			\drawedge(9,4){}

			\node[linecolor=White](M)(10,-1){$\gothF = \{\{a,b\}, \{a,b,c\}, \{a,b,c,d\}\}$}
			
		\end{picture}
		} \hspace{2cm} \subfigure[Zielonka Tree of $\gothF$]{
		\label{subfigure:tree}
			\unitlength = 1.2mm
		\begin{picture}(28,28)(-7,0)
%			\put(-7,0){\framebox(28,28){}}
			\gasset{Nw=3.5,Nh=3.5,AHnb=0}
			\node[Nw=7,Nh=4,Nmr=4](abcd)(6,25){$abcd$}
			\node[Nmr=0,Nw=5,Nh=4](bcd)(-1,18){$bcd$}
			\node[Nmr=0,Nw=5,Nh=4](acd)(6,18){$acd$}
			\node[Nmr=0,Nw=5,Nh=4](abd)(13,18){$abd$}
			\node[Nmr=3,Nw=4](ab)(13,10.5){$ab$}
			\node[Nmr=0](a)(10,4){$a$}
			\node[Nmr=0](b)(16,4){$b$}
			\drawedge(abcd,bcd){}
			\drawedge(abcd,acd){}
			\drawedge(abcd,abd){}
			\drawedge(abd,ab){}
			\drawedge(ab,b){}
			\drawedge(ab,a){}
		\end{picture}
		}
	\end{center}
\caption{Recurring Example}
\label{figure:example}
\end{figure}

\noindent\textbf{Winning Strategies.} A strategy $\sigma$ for Eve is \textit{surely winning} (or \textit{sure)} from a state $s$ for the winning condition $\Phi$ if any play consistent with $\sigma$ belongs to $\Phi$, and \textit{almost-surely winning} (or \textit{almost-sure}) if for any strategy $\tau$ for Adam, $\mathbb{P}_s^{\sigma, \tau}(\Phi) = 1$. The \textit{sure} and \textit{almost-sure} regions are the sets of states from which she has a sure (resp. almost-sure) strategy.

\section{Former results in memory bounds and randomisation}
\label{section:former}

\subsection{Pure strategies}

	There has been intense research since the sixties on the non-stochastic setting, \textit{i.e.} pure strategies and 2-player arenas. B\"uchi and Landweber showed the determinacy of Muller games in \cite{BuchiLandweberAMS69}. Gurevich and Harrington used the LAR (Latest Appearance Record) of McNaughton to prove their \textit{Forgetful Determinacy} theorem \cite{GurevichHarringtonSTOC82}, which shows that a memory of size $|\calC|!$ is sufficient for any game that uses only colours from $\calC$, even when the arena is infinite. This result was later refined by Zielonka in \cite{ZielonkaTCS98}, using a representation of the Muller conditions as trees:

\begin{definition}[Zielonka Tree of a Muller condition]
\label{definition:zielonkatree}
	The Zielonka Tree $\zfc$ of a winning condition $\m \subseteq \mathcal{P}(\calC)$ is defined  inductively as follows:
	\begin{enumerate}
		\item If $\calC \notin \calF$, then $\zfc  = \calZ_{\overline{\calF},\calC}$, where $\overline{\m} = \mathcal{P}(\calC) \setminus \m$.
		\item If $\calC \in \calF$, then the root of $\zfc$ is labelled with $\calC$. Let $\calC_1, \calC_2, \ldots, \calC_k$ be all the maximal sets in $\{U \notin \calF \mid U \subseteq \calC\}$. Then we attach to the root, as its subtrees, the Zielonka trees of $\m \upharpoonright \calC_i$, \textit{i.e.} the $\calZ_{\m \upharpoonright \calC_i,\calC_i}$, for $i = 1 \ldots k$.
	\end{enumerate}
	Hence, the Zielonka tree is a tree with nodes labelled by sets of colours. A node of $\zfc$ is an Eve node if it is labelled with a set from $\calF$, otherwise it is an Adam node.
\end{definition}

	A later analysis of this construction by Dziembowski, Jurdzinski and Walukiewicz in \cite{DziembowskiJurdzinskiWalukiewiczLICS97} led to an optimal and asymmetrical bound on the memory needed by the players to define sure strategies:

\begin{definition}[Number $m_\m$ of a Muller condition]
	Let $\m \subseteq \calP(\calC)$ be a Muller condition, and $\calZ_{\m_1,\calC_1}, \calZ_{\m_2,\calC_2}, \ldots , \calZ_{\m_k,\calC_k}$ be the subtrees attached to the root of the tree $\zfc$. We define the number $m_\m$ inductively as follows: 

	$m_\m = \left\{	\begin{array}{ll}
					1 & \text{if }\zfc \text{ does not have any subtrees,} \\
					\displaystyle{\max \{m_{\m_1}, m_{\m_2}, \ldots , m_{\m_{k}} \}} & \text{if }\calC \notin \m \text{ (Adam node),} \\
					\displaystyle{\sum_{i=1}^{k} m_{\m_i}} & \text{if }\calC \in \m \text{ (Eve node).} \\
				\end{array}\right .$
\end{definition}

\begin{theorem}[\cite{DziembowskiJurdzinskiWalukiewiczLICS97}]
	\label{theorem:pureupper}
	If Eve has a sure strategy in a 2-player Muller game with the winning condition $\calF$, she has a  pure sure strategy with at most $m_{\calF}$ memory states. Furthermore, there is a 2-player arena $\am$ such that Eve has a sure strategy, but none of her sure strategies have less than $\mf$ memory states. \qed
\end{theorem}

\begin{theorem}[\cite{Cha07-FSTTCS}]
	\label{theorem:pureupperrandomised}
	If Eve has an almost-sure strategy in a 2$\frac{1}{2}$-player Muller game with the winning condition $\calF$, she has a pure almost-sure with at most $m_{\calF}$ memory states. \qed
\end{theorem}

\subsection{Memory reduction through randomisation}

Randomised strategies are more general than pure strategies, and in some cases, they are also more compact. In \cite{ChatterjeedeAlfaroHenzingerQEST04}, a first result showed that upward-closed conditions admit memoryless randomised strategies, while they don't admit memoryless pure strategies:

\begin{theorem}[\cite{ChatterjeedeAlfaroHenzingerQEST04}]
\label{theorem:upwardsclosed}
	If Eve has an almost-sure strategy in a 2$\frac{1}{2}$-player Muller game with an upward-closed winning condition, she has a randomised almost-sure strategy. \qed
\end{theorem}

This result was later extended in \cite{ChatterjeeFOSSACS07}, by removing the leaves attached to a node of the Zielonka Tree representing an upward-closed subcondition:

\begin{definition}[\cite{ChatterjeeFOSSACS07}]
\label{definition:chatterjee}
	Let $\calF \subseteq \calP(\calC)$ be a Muller condition, and $\calZ_{\calF_1,\calC_1}, \calZ_{\calF_2,\calC_2}, \ldots ,\linebreak \calZ_{\calF_{k},\calC_{k}}$ be the subtrees attached to the root of the tree $\zfc$. We define the number $m_\m^U$ inductively as follows: 

	$m_\m^U = \left\{	\begin{array}{ll}
					1 & \text{if }\zfc \text{ does not have any subtrees,} \\
					1 & \text{if }\m \text{ is upward-closed,} \\
					\displaystyle{\max \{m^U_{\m_1}, m^U_{\m_2}, \ldots , m^U_{\m_{k}}\}} & \text{if }\calC \notin \m \text{ (Adam node),} \\
					\displaystyle{\sum_{i=1}^{k} m^U_{\m_i}} & \text{if }\calC \in \m \text{ (Eve node).} \\
				\end{array}\right .$
\end{definition}

\begin{theorem}[\cite{ChatterjeeFOSSACS07}]
\label{theorem:chatterjee}
	If Eve has an almost-sure strategy in a 2$\frac{1}{2}$-player Muller game with the winning condition $\calF$, she has a randomised almost-sure strategy with at most $m^U_{\calF}$ memory states. \qed
\end{theorem}

\section{Randomised Upper Bound}
\label{section:upper}

The upper bound of Theorem~\ref{theorem:chatterjee} is not tight for all conditions. For example, the number $m_{\gothF}^U$ of the condition $\gothF$ in Figure~\ref{subfigure:tree} is three, while there is always an almost-sure strategy with two memory states. We present here yet another number for any Muller condition $\calF$, denoted $\rf$, that we compute from the Zielonka Tree:

\begin{definition}[Number $\ram$ of a Muller condition]
\label{definition:rf}
	Let $\m \subseteq \mathcal{P}(\calC)$ be a Muller condition, where the root has $k+l$ children, $l$ of them being leaves. We denote by $\calZ_{\m_1,\calC_1}, \calZ_{\m_2,\calC_2}, \ldots ,\linebreak \calZ_{\m_k,\calC_k}$ the non-leaves subtrees attached to the root of $\zfc$. We define $\rf$ inductively as follows: 

	$\rf =	 \left\{	\begin{array}{ll}
					1 & \text{if }\zfc \text{ does not have any subtrees,} \\
					\displaystyle{\max \{1,r_{\m_1},r_{\m_2},\ldots,r_{\m_k}\}} & \text{if }\calC \notin \m \text{ (Adam node),} \\
					\displaystyle{\sum_{i=1}^k r_{\m_i}} & \text{if }\calC \in \m \text{ (Eve node) and } l=0, \\
					\displaystyle{\sum_{i=1}^k r_{\m_i}} +1 & \text{if }\calC \in \m \text{ (Eve node) and } l>0. \\

				\end{array}\right .$

\end{definition}

The first remark is that if $\emptyset \in \calF$, $\rf$ is equal to $\mf$: as the leaves belong to Eve, the fourth case cannot occur. In the other case, the intuition is that we merge leaves if they are siblings. For example, the number $r_\gothF$ for our recurring example is two: one for the leaves labelled $bcd$ and $acd$, and one for the leaves labelled $a$ and $b$. The number $m_\gothF$ is four (one for each leaf), and $m_\gothF^U$ is three (one for the leaves labelled $a$ and $b$, and one for each other leaf). This section will be devoted to the proof of Theorem~\ref{theorem:upper}:

\begin{theorem}[Randomised upper bound]
\label{theorem:upper}
	If Eve has an almost-sure strategy in a 2-$\frac{1}{2}$ player Muller game with the winning condition $\calF)$, she has an almost-sure strategy with memory $\rf$.
\end{theorem}

Let $\calG = (\calF, \calA)$ be a game defined on the set of colours $\calC$ such that Eve wins from any initial node. We describe in the next three subsections a recursive procedure to compute an almost-sure strategy for Eve with $\rf$ memory states in each non-trivial case in the definition of $\rf$. We use two lemmas --- Lemmas~\ref{lemma:upperAdam} and \ref{lemma:upperEve} --- that derive directly from similar results in \cite{DziembowskiJurdzinskiWalukiewiczLICS97} and \cite{Cha07-FSTTCS}. The application of these principles to the game $\gothG$ in Figure~\ref{figure:example} builds a randomised strategy with two memory states \textit{left} and \textit{right}. In \textit{left}, Eve sends the token to ($\nwarrow$ \textit{or} $\swarrow$) and in \textit{right}, to ($\nearrow$ \textit{or} $\searrow$). The memory switches from \textit{right} to \textit{left} with probability one when the token visits a $c$, and from \textit{left} to \textit{right} with probability $\frac{1}{2}$ at each step.

\subsection{$\calC$ is winning for Adam}
\label{subsection:41}

In the case where Adam wins the set $\calC$, the construction of $\sigma$ relies on Lemma~\ref{lemma:upperAdam}:

\begin{lemma}
\label{lemma:upperAdam}
Let $\calF \subseteq \calP(\calC)$ be a Muller winning condition such that $\calC \notin \calF$, and $\calA$ be a 2$\frac{1}{2}$-player arena such that Eve wins everywhere. There are subarenas $\calA_1 \ldots \calA_n$ such that:
	\begin{itemize}
		\item $i \neq j \Rightarrow \calA_i \cap \calA_j = \emptyset$;
		\item $ \forall i, \calA_i$ is a trap for Adam in the subarena $\calA \setminus \Attr_E \left( \displaystyle{\cup_{j=1}^{i-1}} \calA_j \right)$;
		\item $ \forall i, \chi(\calA_i)$ is included in the label $E_i$ of a child of the root of $\zfc$, and Eve wins everywhere in $(\calA_i, \calF \upharpoonright E_i)$;
		\item $ \calA = \Attr_E(\displaystyle{\cup_{j=1}^n} \calA_j )$.
	\end{itemize}
\end{lemma}

Let the subarenas $\calA_i$ be the ones whose existence is proved in this lemma. We denote by $\sigma_i$ the almost-sure strategy for Eve in $\calA_i$, and by $a_i$ the attractor strategy for Eve to $\calA_i$ in the arena $\calA \setminus \Attr(\displaystyle{\cup_{j=1}^{i-1}} \calA_j)$. We identify the memory states of the $\sigma_i$, so their union has the same cardinal as the largest of them. For a state $s$, if $i = \min \{j \mid s \in \displaystyle{\Attr_E(\cup_{k=1}^j} \calA_k)\}$, we define $\sigma(s,m)$ by:

\begin{itemize}
	\item if $s \in \calA_i$

		\begin{itemize}
			\item $\sigmau(s,m) = \sigmau_i(s, m)$
			\item $\sigman(s,m) = \sigman_i(s,m)$
		\end{itemize}
	\item if $s \in \Attr_E(\cup_{k=1}^i \calA_k) \setminus \calA_i$
		\begin{itemize}
			\item $\sigmau(s,m) = m$
			\item $\sigman(s,m) = a_i(s)$
		\end{itemize}
\end{itemize}

By induction hypothesis over the number of colours, we can assume that the strategies $\sigma_i$ have $r_{\!\calF_i}$ memory states. The strategy $\sigma$ uses $\max \{r_{\!\calF_i}\}$ memory states.

\begin{proposition}
\label{proposition:trapAdam}
	$\mathbb{P}^{\sigma,\tau}_{s_0}(\exists i, \Inf(\rho) \subseteq \calA_i) = 1$.
\end{proposition}

\proof
	The subarenas $\calA_i$ are embedded traps, defined in such a way that the token can escape an $\calA_i$ only by going to the attractor of a smaller one. Eve has thus a positive probability of reaching an $\calA_j$ with $j<i$. Thus, if the token escapes one of the $A_i$ infinitely often, the token has probability one to go to an $\calA_j$ with $j<i$. By argument of minimality, after a finite prefix, the token will stay in one of the traps forever.
\qed

The strategy $\sigma_i$ is almost-sure from any state in $\calA_i$. As Muller conditions are prefix-independent, it follows from Proposition~\ref{proposition:trapAdam} that $\sigma$ is also almost-sure from any state in $\calA$.

\subsection{$\calC$ is winning for Eve, and the root of $\zfc$ has no leaves among its children.}
\label{subsection:42}
In this case, the construction relies on the following lemma:

\begin{lemma}
\label{lemma:upperEve}
	Let $\calF \subseteq \calP(\calC)$ be a Muller winning condition such that $\calC \in \calF$, $\calA$ a 2$\frac{1}{2}$-player arena coloured by $\calC$ such that Eve wins everywhere, and $A_i$ the label of a child of the root in $\zfc$. Then, Eve wins everywhere on the subarena $\calA \setminus \Attr_E(\chi^{-1}(\calC \setminus A_i))$ with the condition $\calF \upharpoonright A_i$.
\end{lemma}

Eve has a strategy $\sigma_i$ that is almost-sure from each state in $\calA \setminus \Attr_E(\chi^{-1}(\calC \setminus A_i))$. In this case, the set of memory states of $\sigma$ is $M = \displaystyle{\cup_{i=1}^{k}} (i \times M^i)$. The ``next-move'' and ``memory-update'' functions $\sigman$ and $\sigmau$ for a memory state $m = (i,m^i)$ are defined below:

\begin{itemize}
	\item if $s \in \chi^{-1}(\calC \setminus A_i)$
		\begin{itemize}
			\item $\sigmau(s,(i,m^i)) = (i+1 , m^{i+1})$ where $m^{i+1}$ is any state in $M^{i+1}$
			\item if $s \in S_E$, $\sigman(s,(i,m^i))$ is any successor of $s$ in $\calA$
		\end{itemize}
	\item if $s \in \Attr_E(\chi^{-1}(\calC \setminus A_i))$
		\begin{itemize}
			\item $\sigmau(s,(i,m^i)) = (i, m^i)$
			\item $\sigman(s,(i,m^i)) = a_i(s)$
		\end{itemize}
	\item if $s \in \calA \setminus \Attr_E(\chi^{-1}(\calC \setminus A_i))$
		\begin{itemize}
			\item $\sigmau(s,(i,m^i)) = (i, \sigmau_i(s,m^i))$
			\item $\sigman(s,(i,m^i)) = \sigman_i(s,m^i)$
		\end{itemize}
\end{itemize}

Once again, we can assume that the memory $M_i$ of the strategy $\sigma_i$ is of size $r_{\calF \upharpoonright A_i}$. Here, however, the memory set of $\sigma$ is the disjoint union of the $M_i$', so $\sigma$'s needs the sum of the $\{r_{\calF \upharpoonright A_i}\}$'s.

\begin{proposition}
\label{proposition:constant}
	Let $\mathfrak{uc}$ be the event ``the top-level memory of $\sigma$ is ultimately constant''. Then, $\mathbb{P}^{\sigma,\tau}_{s_0}(\rho \in \Phi_{\calF} \mid \mathfrak{uc}) = 1$.
\end{proposition}

\proof
	We call $i$ the value of the top-level memory at the limit. After a finite prefix, the token stops visiting $\chi^{-1}(\calC \setminus A_i)$. Thus, with probability one, it also stops visiting $\Attr_E(\chi^{-1}(\calC \setminus A_i))$. From this point on, the token stays in the arena $\calA_i$, where Eve plays with the almost-sure strategy $\sigma_i$. Thus, $\mathbb{P}^{\sigma,\tau}(\rho \in \Phi_{\calF\upharpoonright A_i} \mid \mathfrak{uc}) = 1$, and, as $\Phi_{\calF\upharpoonright A_i} \subseteq \Phi_{\calF}$, Proposition~\ref{proposition:constant} follows.
\qed

\begin{proposition}
\label{proposition:variation}
	If the top-level memory takes each value in $1\ldots k$ infinitely often, then surely, $\forall i \in 1\ldots k, \chi(\Inf(\rho)) \nsubseteq A_i$.
\end{proposition}

\proof
	The update on the top-level memory follows a cycle on $1\ldots k$, leaving $i$ only when the token visits $\chi^{-1}(\calC \setminus A_i)$. Thus, in order for the top-level memory to change continuously, the token has to visit each of the $\chi^{-1}(\calC \setminus A_i)$ infinitely often. Proposition \ref{proposition:variation} follows.
\qed

\subsection{$\calC$ is winning for Eve, and the root of $\zfc$ has at least one leaf in its children.}
\label{subsection:43}

As in the previous section, the construction relies on Lemma~\ref{lemma:upperEve}. In fact, the construction for children which are not leaves, labelled $A_1,\ldots,A_k$, is exactly the same. The difference is that we add here a single memory state ---$0$--- that represents all the leaves (labelled $A_{-1},\ldots,A_{-l}$). The memory states are thus updated modulo $k+1$, and not modulo $k$. The ``next-move'' function of $\sigma$ when the top-level memory is $0$ is an even distribution over all the successors in $A$ of the current state. The ``memory-update'' function has probability $\frac{1}{2}$ to stay into $0$, and $\frac{1}{2}$ to go to $(1,m_1)$, for some memory state $m_1 \in M_1$. Thus, $\sigma$ uses memory $\sum_{i=1}^k r_{\calF_i} + 1$. We prove now that $\sigma$ is almost-sure. The structure of the proof is the same as in the former section, with some extra considerations for the memory state $0$.

\begin{proposition}
	\label{proposition:constantbis}
	Let $\mathfrak{uc}$ be the event ``the top-level memory of $\sigma$ is ultimately constant and different from 0''. Then, $\mathbb{P}^{\sigma,\tau}_s(\rho \in \Phi_{\calF} \mid \mathfrak{uc}) = 1$.
\end{proposition}

\proof
	The proof is exactly the same as the one of Proposition~\ref{proposition:constant}.
\qed

\begin{proposition}
	\label{proposition:constant0}
	 The event ``the top-level memory is ultimately constant and equal to 0'' has probability 0.
\end{proposition}

\proof
	When the top-level memory is 0, the memory-update function has probability $\frac{1}{2}$ at each step to switch to 1. Proposition~\ref{proposition:constant0} follows.
\qed

Proposition~\ref{proposition:variationbis} considers the case where the top-level memory evolves continuously. By definition of the memory update, this can happen only if all the memory states are visited infinitely often.

\begin{proposition}
	\label{proposition:variationbis}
	Let $\mathfrak{ec}$ be the event ``the top-level memory takes each value in $0\ldots k$ infinitely often''. Then, $\forall i\in -l \ldots k, \mathbb{P}^{\sigma,\tau}_s(\chi(\Inf(\rho)) \subseteq \calC_i \mid \mathfrak{ec}) = 0$.
\end{proposition}
	
\proof
	As in the proof of Proposition~\ref{proposition:variation}, from the fact that the memory is equal to each of the $i \in 1\ldots k$ infinitely often, we can deduce that the token surely visits each of the $\calC \setminus A_i$ infinitely often. We only need to show that, with probability one and for any $j\in 1 \ldots l$, the set of limit states is not included in $A_{-j}$. The Zielonka Trees of the conditions $\calF\upharpoonright A_{-j}$ are leaves. This means that they are trivial conditions, where all the plays are winning for Adam. Consequently, in this case, Lemma~\ref{lemma:upperEve} guarantees that $\Attr_E(\chi^{-1}(\calC \setminus A_{-j}))$ is the whole arena. The definition of $\sigma$ in the memory state $(0)$ is to play legal moves at random. There is thus a positive probability that Eve will play according to the attractor strategy $a_j$ long enough to guarantee a positive probability that the token visits $\chi^{-1}(\calC \setminus A_{-j})$. To be precise, for any $s\in S$, this probability is greater than $(2 \cdot |S|)^{-|S|}$. Thus, with probability one, the token visits each $\chi^{-1}(\calC \setminus A_{-j})$ infinitely often. Proposition \ref{proposition:variationbis} follows.
\qed

The initial case, where the Zielonka tree is reduced to a leaf, is trivial: the winner does not depend on the play. Thus, Theorem~\ref{theorem:upper} follows from Sections~\ref{subsection:41},~\ref{subsection:42}, and~\ref{subsection:43}.

\section{Lower Bound}
\label{section:lower}
In this section, we consider lower bounds on memory, \textit{i.e.} if we fix a Muller condition $\calF$ on a set of colours $\calC$, the minimal size of the memory set that is enough to define randomised almost-sure strategies for Eve on any arena coloured by the set $\calC$. In his thesis, Majumdar showed the following theorem:

\begin{theorem}[\cite{Majumdar03}]
\label{theorem:globallowerbound}
	For any set of colours $\calC$, there is a 2-player Muller game $\calG_\calC = (\calA_\calC,\calF_\calC)$ such that Eve has an almost sure, but none of her almost-sure strategies have less than $\frac{|\calC|}{2}!$ memory states.
\end{theorem}

However, this is a general lower bound on \textit{all} Muller conditions, while we aim to find specific lower bounds for \textit{each} condition. We prove here that there is a lower bound for each Muller condition that matches the upper bound of Theorem~\ref{theorem:upper}:

	\begin{theorem}
	\label{theorem:lower}
		Let $\calF$ be a Muller condition on $\calC$. There is a 2-player arena $\calA_\calF$ over $\calC$ such that Eve has a sure strategy, but none of her almost-sure strategies have less than $\rf$ memory states.
	\end{theorem}

	As the construction of the upper bound was based on the Zielonka tree, the lower bound is based on the Zielonka \textit{DAG}:

	\begin{definition}
	\label{definition:dag}
		The Zielonka DAG $\dfc$ of a winning condition $\m \subseteq \mathcal{P}(\calC)$ is derived from $\zfc$ by merging the nodes which share the same label.
	\end{definition}

	\subsection{Cropped DAGs}
	\label{subsection:croppedDAG}
	
	The relation between $\rf$ and the shape of $\dfc$ is asymmetrical: it depends directly on the number of children of Eve's nodes, and not at all on the number of children of Adam's nodes. The notion of \textit{cropped DAG} is the next logical step: a sub-DAG where Eve's nodes keep all their children, while each node of Adam keeps only one child:

		\begin{definition}
		\label{definition:croppedDAG}
			A DAG $\calE$ is a cropped DAG of a Zielonka DAG $\dfc$ \iff
			\begin{itemize}
				\item The nodes of $\calE$ are nodes of $\dfc$, with the same owner and label.
				\item There is only one node without predecessor in $\calE$, which we call the root of $\calE$. It is the root of $\dfc$, if it belongs to Eve; otherwise, it is one of its children.
				\item The children of a node of Eve in $\calE$ are exactly its children in $\dfc$.
				\item A node of Adam has exactly one child in $\calE$, chosen among his children in $\dfc$, provided there is one. If it has no children in $\dfc$, it has no children in $\calE$.
			\end{itemize}
		\end{definition}
		
		Cropped DAG resemble Zielonka DAGs: the nodes belong to either Eve or Adam, and they are labelled by sets of states. We can thus compute the number $r_\calE$ of a cropped DAG $\calE$ in a natural way. In fact, this number has a more intuitive meaning in the case of cropped DAGs: if the leaves belong to Eve, it is the number of branches; if Adam owns the leaves, it is the number of branches with the leaf removed. Furthermore, there is a direct link between the cropped DAGs of a Zielonka DAG $\dfc$ and the number $\rf$:
		
		\begin{proposition}
		\label{proposition:memfrerf}
			Let $\calF$ be a Muller condition on $\calC$, and $\dfc$ be its Zielonka DAG. Then there is a cropped DAG $\calE^*$ such that $r_{\calE^*} = \rf$. \qed
		\end{proposition}
		
	\subsection{From cropped DAGs to arenas}
	\label{subsection:DAGarenas}
	
		From any cropped DAG $\calE$ of $\dfc$, we define an arena $\are$ which follows roughly the structure of $\calE$: the token starts from the root, goes towards the leaves, and then restarts from the root. In her nodes, Eve can choose to which child she wants to go. Adam's choices, on the other hand, consists in either stopping the current traversal or allowing it to proceed.
	
		We first present two ``macros'', depending on a subset of $\calC$:
		\begin{itemize}
			\item in $\Pick^*(C)$, Adam can visit any subset of colours in $C$;
			\item in $\Pick(D)$, he must visit exactly one colour in $D$.
		\end{itemize}
		Both are represented in Figure~\ref{figure:pick}, and they are the only occasions where colours are visited in $\are$: all the other states are colourless.

\begin{figure}[ht]
	\begin{center}
		\unitlength = 1.5mm
		\subfigure[$\Pick ^* (C)$]{
		\label{subfigure:DAGarenapick*}
			\begin{picture}(25,14)
%			\put(0,0){\framebox(27,14){}}
			\gasset{Nw=3,Nh=3}
			\node(c1)(4,9){$c_1$}
			\node(c1bis)(4,5){}
			\node(ck)(14,9){$c_i$}
			\node(ckbis)(14,5){}
			\node(cz)(24,9){$c_k$}
			\node(czbis)(24,5){}
			\node[Nw=2.6,Nh=2.6,Nmr=0,dash={.15 .3}0](dots1)(9,7){$\cdots$}
			\node[Nw=2.6,Nh=2.6,Nmr=0,dash={.15 .3}0](dots2)(19,7){$\cdots$}
			\node[Nw=2.6,Nh=2.6,iangle=90,Nmarks=i,ilength=2,Nmr=0](top)(1,13){}
			\node[Nw=2.6,Nh=2.6,fangle=270,Nmarks=f,flength=2,Nmr=0](bottom)(27,1){}

			\drawedge(top,c1){}
			\drawedge(c1,dots1){}
			\drawedge(dots1,ck){}
			\drawedge(ck,dots2){}
			\drawedge(dots2,cz){}
			\drawedge[curvedepth=2](cz,bottom){}

			\drawedge[curvedepth=-2](top,c1bis){}
			\drawedge(c1bis,dots1){}
			\drawedge(dots1,ckbis){}
			\drawedge(ckbis,dots2){}
			\drawedge(dots2,czbis){}
			\drawedge(czbis,bottom){}

			\node[linecolor=White](pipal)(9,1.5){$C = \{c_1 \ldots c_k\}$}

			\end{picture}
		}\hfil\subfigure[$\Pick (D)$]{
		\label{subfigure:DAGarenapick}
			\begin{picture}(25,14)
%			\put(2.5,0){\framebox(23,14){}}
			\gasset{Nw=3,Nh=3}
			\node(d1)(6,7){$d_1$}
			\node(dh)(14,7){$d_i$}
			\node(dt)(22,7){$d_k$}
			\node[linecolor=White](dots1)(10,7){$\cdots$}
			\node[linecolor=White](dots2)(18,7){$\cdots$}
			\node[Nw=2.6,Nh=2.6,iangle=90,Nmarks=i,ilength=2,Nmr=0](top)(09,11){}
			\node[Nw=2.6,Nh=2.6,fangle=270,Nmarks=f,flength=2,Nmr=0](bottom)(19,3){}

			\drawedge(top,d1){}
			\drawedge(top,dh){}
			\drawedge(top,dt){}
			\drawedge(d1,bottom){}
			\drawedge(dh,bottom){}
			\drawedge(dt,bottom){}

			\node[linecolor=White](pipal)(9,1.5){$D = \{d_1 \ldots d_k\}$}

			\end{picture}
		}
	\end{center}
	\caption{$\Pick^*(C)$ and $\Pick(D)$}
	\label{figure:pick}
\end{figure}

	Eve's states in the arena $\are$ are in bijection with her nodes in $\calE$. Adam's nodes, on the other hand, are in bijection with the pairs parent-child of $\calE$, where the parent belongs to Eve and the child to Adam. 

	In the state corresponding to the node $n$, Eve can send the token to any state of the form $n-c$. In states corresponding to leaves, Eve has no decision to take, and Adam can visit any colours in the label of the leaf ($\Picks$ procedure). The token is then sent back to the root.

	Adam's moves do not involve the choice of a child: by Definition~\ref{definition:croppedDAG}, Adam's nodes in $\calE$ have but one child. Instead, he can either stop the current traversal, or, if the current node is not a leaf, allow it to proceed to its only child. If he chooses to stop, Adam has to visit some coloured states before the token is sent back to the root. The available choices depend on the labels of both the current and the \textit{former} nodes --- which is why there are as many copies of Adam's nodes in $\are$ as they have parents in $\calE$. If the parent is labelled by $E$, and the current node by $A$, the token goes through $\Picks(E)$ and $\Pick(E \setminus A)$. Adam can thus choose any number of colours in $E$, as long as he chooses at least one outside of $A$.

\begin{figure}[ht]
 \unitlength = 2mm
	\begin{center}
		\subfigure[Edge ``$E$'' - ``$A$'' when ``$A$'' is a node]{
		\label{subfigure:Anode}
			\begin{picture}(31,13)(-6,-1)
%			\put(0,0){\framebox(19,12){}}
			\scriptsize
			\gasset{Nw=2,Nh=2}
			
			\node(top)(1,9){$E$}
			\imark[iangle=90](top)
			\node[Nmr=0](node)(1,5){$A$}
			\node(bottom)(1,1){$E'$}
			\drawedge(top,node){}
			\drawedge(node,bottom){}

			\node(topdag)(7,9){}
			\imark[iangle=90](topdag)
		 	\nodelabel[ExtNL=y,NLdist=.27,NLangle=320](topdag){$E$}
			\node[Nmr=0](nodedag)(7,5){}
		 	\nodelabel[ExtNL=y,NLdist=.27,NLangle=320](nodedag){$E-A$}
			\node(bottomdag)(7,1){}
		 	\nodelabel[ExtNL=y,NLdist=.27,NLangle=320](bottomdag){$E'$}
			\drawedge(topdag,nodedag){}
			\drawedge(nodedag,bottomdag){}
			
			\node[Nadjust=wh,Nmr=0,dash={.5}0](pick*)(15,9){$\Pick^* (E)$}
			\imark[iangle=90](pick*)
			\node[Nadjust=wh,Nmr=0,dash={.5}0](pick)(15,5){$\Pick (E \setminus A)$}
			\node[linecolor=White](root)(15,1){\texttt{root}}
			\drawedge(pick*,pick){}
 			\drawedge(pick,root){}
 			
 			\drawbpedge(nodedag,0,8,pick*,180,10){}

			\end{picture}
		}
		\hfil
		\subfigure[Edge ``$E$'' - ``$A$'' when ``$A$'' is a leaf]{
		\label{subfigure:Aleaf}
			\begin{picture}(25,13)(-5,-1)
%			\put(-2,0){\framebox(19,12){}}
			\scriptsize
			\gasset{Nw=2,Nh=2}
			
			\node(top)(1,7){$E$}
			\imark[iangle=90](top)
			\node[Nmr=0](leaf)(1,3){$A$}
			\rmark(leaf)
			\drawedge(top,leaf){}

			\node[Nadjust=wh,Nmr=0,dash={.5}0](pick*)(11,9){$\Pick^* (E)$}
			\imark[iangle=90](pick*)
			\node[Nadjust=wh,Nmr=0,dash={.5}0](pick)(11,5){$\Pick (E \setminus A)$}
			\node[linecolor=White](root)(11,1){\texttt{root}}
			\drawedge(pick*,pick){}
 			\drawedge(pick,root){}

			\end{picture}
		}
	\end{center}
	\caption{Adam's states in $\are$.}
	\label{figure:Achoice}
\end{figure}
		
	\subsection{Winning strategy and branch strategies}
	\label{subsection:DAGstrategies}
		
		We first describe a sure strategy $\varsigma$ for Eve in the game $(\are,\muller)$. Its memory states are the branches of $\calE$, and do not change during a traversal. If the current memory state is $b = E_1 A_1 \ldots E_\ell (A_\ell)$, Eve's moves follow the branch $b$: in $E_i$, she goes to $E_i - A_i$. When Adam stops the traversal at the $i$th step, Eve updates her memory as follows:

		\begin{itemize}
			\item	If $E_i$ has zero or one child in $\calE$, the memory is unchanged;
			\item	otherwise, the new memory branch has $E_1 A_1 \ldots E_i A$ as a prefix, where $A$ is the next child of $E_i$, or the first one if $A_i$ was the last.
		\end{itemize}
	
		\begin{proposition}
		\label{proposition:Evewinsare}
			The strategy $\varsigma$ is surely winning for Eve in the game $(\are,\muller)$.
		\end{proposition}
	
		\proof
			Let $\rho$ be a play consistent with $\varsigma$. We denote by $i$ the smallest integer such that traversals stops infinitely often at the $i$th step. After a finite prefix, the first $2i-1$ nodes in the memory branch are constant, and we denote them by $E_1 A_1 E_2 \ldots E_i$. From this point on, the colours visited belong to $E_i$. Furthermore, each time a traversal stops at step $i$, a state is visited outside of the current $A_i$, which changes afterwards to the next, in a circular way. It follows that $\Inf(\rho) \subseteq E_i$, and, for any child $A$ of $E_i$ in $\calE$, $\Inf(\rho) \nsubseteq A$. Thus $\rho$ is winning for Eve. Proposition~\ref{proposition:Evewinsare} follows.
		\qed

		\medskip

		Obviously, Adam has no winning strategy in $\are$. However, we describe the class of \textit{branch strategies}, whose point is to punish any attempt of Eve to win with less than $\rf$ memory states. There is one such strategy $\tau_b$ for each branch $b$ in $\calE$ (whence the name), and the principle is that $\tau_b$ stops the traversal as soon as Eve deviates from $b$:
	
		\begin{definition}
			The branch strategy $\tau_b$ for Adam in $\are$, corresponding to the branch $b = E_1 A_1 E_2 \ldots E_\ell (A_\ell)$ in $\calE$, is a positional strategy whose moves are described below.
			
			\begin{itemize}
				\item In a state $E-A$ such that $\exists i, E = E_i \wedge A \neq A_i$: stop the traversal and visit $A_i$;
				\item in a state $E-A$ such that $\exists i, E = E_i \wedge A = A_i$: send the token to $E_{i+1}$;
				\item in the state $E_\ell-A_\ell$, or the leaf $E_\ell$: visit the colours of $E_\ell$.
			\end{itemize}
		\end{definition}

		No move is given for a state $E-A$ such that $\forall i, E \neq E_i$, as these states are not reachable from the root when Adam plays $\tau_b$. Notice also that when Adam chooses to stop a traversal in a state $E_i - A$, he \textit{can} visit exactly the colours of $A_i$: as $A$ and $A_i$ are maximal subsets of $E_i$, there is at least one state in $A_i \setminus A$ that he can pick in the $\Pick(E_i \setminus A)$ area.
	
	\subsection{Winning against branch strategies}
	\label{subsection:DAGbounds}
	
	The key idea of the proof of Theorem~\ref{theorem:lower} is that if two branches $b$ and $b'$ of $\calE$ are too different, Eve needs different memory states to win against $\tau_b$ and $\tau_{b'}$.

		\begin{proposition}
		\label{proposition:randomare}
			Let $\sigma = (M,\sigman,\sigmau)$ be an almost-sure strategy for Eve in $(\are,\muller)$. Then $\sigma$ has memory at least $\re$.
		\end{proposition}

		\proof
			Let $b = E_1 A_1 \ldots E_\ell (A_\ell)$ be a branch of $\calE$ and $\tau_b$ be the corresponding branch strategy for Adam. By definition of $\tau_b$, the set of colours visited in a traversal consistent with $\tau_b$ is one of the $A_i$'s, or $E_\ell$ \iff Eve plays along $b$. As $\sigma$ is almost-sure, there must be a memory state $m$ such that Eve has a positive probability to play along $b$. It is also necessary to ensure that none of the $A_i$'s is visited infinitely often, with the possible exception of $A_\ell$. So, if Eve has a positive to play along a branch $b'$ when she is in the memory state $m$, $E_1 A_1 \ldots E_\ell$ must be a prefix of $b'$. It follows that a single memory state can be suitable against two strategies $\tau_b$ and $\tau_{b'}$  with $b = E_1 A_1 \ldots E_\ell (A_\ell)$ and $b' = E'_1 A'_1 \ldots E'_{\ell'} (A'_{\ell'})$ only if $\ell = \ell'$ and $\forall i \le \ell, E_i = E'_i$. By Definition~\ref{definition:rf}, the underlying equivalence relation has $\re$ equivalence classes. Proposition~\ref{proposition:randomare} follows.
		\qed
		
		By Proposition~\ref{proposition:memfrerf}, there is a cropped DAG $\calE$ of $\dfc$ such that $\re = \rf$. So, in general, Eve needs randomised strategies with memory $\rf$ in order to win games whose winning condition is $\calF$. This completes the proof of Theorem~\ref{theorem:lower}.
	
\section{Conclusion}
\label{section:conclusions}

We have provided better and tight bounds for the memory needed to define almost sure winning randomised strategies. This allows us to characterise the class of Muller conditions which admit randomised memoryless strategies:

\begin{corollary}
Eve admits randomised memoryless almost-sure strategies for a Muller condition $\calF$ if and only if all her nodes in $\zfc$ have either one child, or only leave children. \qed
\end{corollary}

This yields a {\tt NP} algorithm for the winner problem of such games, as solving 1$\frac{1}{2}$-player Muller games is {\sc Ptime} \cite{ChatterjeedeAlfaroHenzingerQEST04}. Another consequence of our result is that for each Muller condition, at least one of the players cannot improve its memory through randomisation:

\begin{corollary}
Let $\calF$ be a Muller condition. If $\emptyset \in \calF$, Eve needs as much memory for randomised strategy as for pure strategies. Otherwise, Adam does. \qed
\end{corollary}

Our proof of lower bound also improves on the size of the witness arena: it is roughly equivalent to the size of the Zielonka DAG, instead of the size of the Zielonka tree. Whether these bounds still hold for arenas of size polynomial in the number of colours remains an open question, except for special cases like Streett games \cite{HornIPL07}.

\end{document}